\documentclass[a4paper, 10pt, conference]{ieeeconf}      

\usepackage[all]{xy}
\usepackage{color}
\usepackage{bm}
\usepackage{graphicx}
\usepackage{color}
\usepackage{wrapfig}
\usepackage{times,mathptm}
\usepackage{url}
\usepackage{amssymb, amsmath}

\IEEEoverridecommandlockouts                              
\title{\LARGE \bf
Smart Finite State Devices: \\ A Modeling Framework for Demand Response Technologies
}

\author{Konstantin Turitsyn, Scott Backhaus, Maxim  Ananyev and Michael Chertkov
\thanks{The work of MC at LANL was carried out under the auspices of the
National Nuclear Security Administration of the U.S. Department of
Energy at Los Alamos National Laboratory under Contract No.
DE-AC52-06NA25396.}
\thanks{K. Turitsyn is with MIT, Mechanical Engineering, Cambridge, MA 02139 {\tt\small turitsyn@mit.edu}}
\thanks{S. Backhaus is with MPA Division at LANL, Los Alamos, NM 87545 {\tt\small backhaus@lanl.gov}}
\thanks{M. Ananyev is with New Economic School, Moscow, Russia {\tt\small maksim.ananjev@gmail.com}}
\thanks{M. Chertkov is with Theory Division \& Center for Nonlinear Studies at LANL,
Los Alamos, NM 87545 and also with New Mexico Consortium, Los Alamos, NM 87544 {\tt\small chertkov@lanl.gov}}
}

\begin{document}
\maketitle
\thispagestyle{empty}
\pagestyle{empty}

\begin{abstract}
We introduce and analyze Markov Decision Process (MDP) machines to model individual devices which are expected to participate in future demand-response markets on distribution grids. We differentiate devices into the following four types: (a) optional loads that can be shed, e.g. light dimming; (b) deferrable loads that can be delayed, e.g. dishwashers; (c) controllable loads with inertia, e.g. thermostatically-controlled loads, whose task is to maintain an auxiliary characteristic (temperature) within pre-defined margins; and (d) storage devices that can alternate between charging and generating. Our analysis of the devices seeks to find their optimal price-taking control strategy under a given stochastic model of the distribution market.
\end{abstract}

\section{Introduction}
\label{sec:Intro}
Automated demand response is often used to manage electrical load during critical system peaks\cite{07LBNL-DR-report,DOE-DR06}.  During a typical event as the system approaches peak load, signaling from the utility results in automated customer load curtailment for a given period of time to avoid overstressing the grid.  Although this type of load control is useful for maintaining system security, automated demand response must evolve further to meet the coming challenge of integrating time-intermittent renewables such as wind or photovoltaic generation.  When these resources achieve high penetration and their temporal fluctuations exceed a level that can be economically mitigated by the remaining flexible traditional generation (e.g. combustion gas turbines), automated demand response will play a large role in maintaining the balance between generation and load.  To fill this role, automated demand response must go beyond today's peak-shaving capability

To follow intermittent generation, automated demand response must be bi-directional control, i.e. it should provide for controlled increases and decreases in load.  The response must also be predictable and preferably non-hysteretic, otherwise the load-generation imbalance may actually be exacerbated.  Predictability would be highly valued by third party companies that aggregate loads into a pool of demand response resources.  Finally, whatever control methodology is implemented, it must also be stable and not exhibit temporal oscillations.  There are several factors that make achieving these demand response goals challenging: the different options for demand response signal, the uncertainty of the aggregate response to that signal, and the inhomogeneity of the underlying ensemble of loads.

The demand response control signal could take several forms: direct load control where some number of loads could be disabled via a utility-controlled switch\cite{92SS,06BKT}; end-use parameter control where an ensemble of loads can be controlled by modifying the set point of the end-use controller, e.g. a thermostat temperature set point\cite{09Cal,11CH,10KSBH}; or indirect control via energy pricing in either a price taking (open loop) or auction (closed loop)\cite{Olympic} setting.  Today's automated demand response for peak-shaving is a form of direct load control which could be adapted and refined for the type of operation we desire, however, it is difficult to assess the impact of demand response on the end user because loads are simply disabled and re-enabled with little concern for the current state of the end use.  Direct control is feasible for a relatively small number of large loads because the communication overhead is not extreme. Individual direct control of a large number of small loads would potentially overburden a communication system, however, ``ensemble'' control using a single parameter for control has been proposed, e.g. set point control for thermostatic loads\cite{09Cal,11CH,10KSBH} and connection rate control for electric vehicle charging\cite{10TSBC}.  However, in these control models, the underlying loads are assumed to be homogeneous (all of the same type), which is advantageous because it allows for a quantifiable measure of the end use impacts and customer discomfort, e.g. increasing all cooling
thermostat set points by 1$^oF$ will generate a decrease in load with a known end-use impact.

To control a large ensemble of {\it inhomogenous} loads with a single demand response signal requires a quantity that applies to all loads, i.e. energy pricing \cite{80STKOPC}.  When given access to energy prices, consumers (or automated controllers acting on their behalf) can make their own local decisions about whether to consume or not.  These local decisions open up new possibilities and also create problems.   The customer is now enabled to automatically modify and perhaps optimize his consumption of energy to maximize his own welfare, which is a combination of his total energy costs and the completion of the load's end use function.  However, without an understanding of how consumers respond to energy prices, the fidelity of the control allowed by the direct or ensemble control schemes described above is lost.  Retail-level double auction markets\cite{Olympic} are an effective way of making demand response via pricing a closed-loop control system, however, a logical outcome of these markets would be locational prices potentially driven distribution system constraints making the regulatory implementation troublesome.  In contrast, a model where retail customers are price takers may avoid some regulatory issues, however, price taking is in essence a form of open loop control which then requires an understanding of how the aggregate load on the system will respond to price.

Our goal in this initial work is to layout the computational framework for discovering the end-use response to these price-taking ``open-loop'' control systems.  We develop state models for several different loads and subject them to a stochastic price signal that represents how energy prices might behave in an grid with a large amount of time-intermittent generation.  We analyze the response of these smart loads using a Markov Decision Process (MDP) to optimize the welfare of the end user. Human owners of the devices have the ability to program the devices in accordance to their strategies and preferences, for instance by adjusting their willingness to sacrifice comfort in exchange for savings on electricity costs. Otherwise, most of the time we assume that the devices operate automatically in accordance to some optimal algorithm that was either preprogrammed by their owners, discovered via adaptive learning\cite{10OLGM}, or programmed by a third-party aggregator.  The resulting load end-use policies can then be turned around to predict the effect of a change in prices on electrical load.  Our long term strategic intention is to analyze the aggregated network effect on power flows of many independent customers and design optimal strategies for both consumers and the power operator.  However, the prime focus of our first publication on the subject is less ambitious. We focus here on description of different load models and analyze the optimal behavior of individual consumers.

The material in the manuscript is organized as follows. We formulate our main assumptions and introduce the general MDP framework in Section \ref{sec:setting}.  Models of four different devices (optional, deferable and control loads and storage devices) are introduced in
Section \ref{sec:devices}.  Our enabling simulation example of a control load (smart thermostat) is presented in Section \ref{sec:Simulations}. We summarize our main results and discuss a path forward in Section \ref{sec:Path}.

\section{Setting the Problems}
\label{sec:setting}

\subsection{Basic Assumptions}
\label{subsec:Basic Assumptions}

Future distribution networks are expected to show complex, collective behavior originating from competitive interaction of individual players of the following three types:
\begin{itemize}
\item Market operator, having full or partial control over the signals sent to devices/customers. The most direct signal is energy price. The operator may also provide subsidies and incentives or impose penalties, however in this manuscript, we will mainly focus on direct price control.

\item Human customers/owners,  who are able to reprogram smart-devices or override their actions.

\item Smart devices, capable of making decisions about their operations. The devices are semi-automatic, i.e. pre-programmed to respond to the signal on a short time scale (measured in seconds-to-minutes) in a specific way,  however the owner of the device may also choose to change the strategy on a longer time-scale (days or weeks). We model the smart devices as finite state machines using a Markov Decision Process (MDP) framework. At the beginning of each interval, a device decides how to change its state based on the current price. Each change comes with a reward expressing actual transactions between the provider and the consumer and the level of consumer satisfaction with the decision. We assume that smart devices are selfish and not collaborative, each optimizing its own reward.
\end{itemize}
In this manuscript we restrict our attention to a simple price-taking strategy of consumer behavior, deferring analysis of more elaborate game-theoretic interactions between the operator and the individual customers to further publications.

We model the external states (that include electricity price, weather, and human behavior) as a stochastic, Markov Chain process, $\{s^{(e)}(t)\}$. At the beginning of the time interval, $t$, the variable describing these factors is set to $s^{(e)}_t$ and changes during the next time step to $s^{(e)}_{t+1}$ with the transition probability $T(s^{(e)}_{t+1}|s^{(e)}_t)$.  The transition probabilities are assumed to be known to the device and statistically stationary, i.e. independent of $t$. (The later assumption can be easily relaxed to account for natural cycles and various external factors.) The probability, $p(s^{(e)};t)$, to observe the external state, $s^{(e)}(t)=s^{(e)}$, at the time $t$,  follows the standard Markov chain equation
\begin{eqnarray} \label{external-chain}
p(s^{(e)};t+1)=\sum_{s_{t}^{(e)}} T(s^{(e)},s_{t}^{(e)}) p(s_t^{(e)}|t).
\label{P-eq}
\end{eqnarray}
We also assume that the Markov chain (\ref{external-chain}) is ergodic and converges after a finite transient to the statistically stationary distribution: $p(s^{(e)};t+1)=p(s^{(e)};t)=p(s^{(e)})$. In the simulation tests that follow we will restrict ourselves to $s^{(e)}$ drawn from a finite set $S^{(e)}$.

\subsection{General Markov Decision Process Framework}
\label{subsec:FSM}

Here we adopt the standard (Markov Decision Process) MDP approach \cite{57Bel,05Put,MDP-package} to the problem of interest: description of smart devices responding to the external (exogenous) Markov process $\{s^{(e)}(t)\}$. MDPs provide a mathematical framework for modeling decision-making in situations where outcomes are partly random and partly under the control of a decision maker. Formally,  the MDP is a 4-tuple, $(S,A,P(\cdot,\cdot),R(\cdot,\cdot))$,  where
\begin{itemize}
\item $S$ is the finite set of states, in our case a direct product of the machine states set $S^{(m)}$, and the externality state set $S^{(e)}$, $S=S^{(m)}\otimes S^{(e)}$.
\item $A$ is a finite set of actions. $A_s$ is the finite set of actions available from state $s\in S$. Within our framework we model only the decisions made by the machine, so the set $A$ consists only of actions associated with the machine, $A = A^{(m)}$.
\item $P_a(s,s')= \Pr(s_{t+1}=s' \mid s_t = s,\, a_t=a)$ is the probability that action $a$ chosen while in state $s=(s^{(m)},s^{(e)})$ at time $t$ will lead to state $s'$ at time $t + 1$. The probabilistic description of the transition allows to account for stochastic nature of the price fluctuations as well as for the randomness in the dynamics of the smart devices.
\item $R_a(s,s')$ is the reward associated with the transition $s \to s'$ if the action $a$ was chosen. In our models, the reward will reflect the price paid for electricity consumption associated with the transition as well as the level of discomfort related to the event.
\end{itemize}
In the most simple setting analyzed in this work, the behavior of the device is modeled via the policy function $\pi(s) : S\to A$ that determines the action chosen by the device for a given state: $a_t=\pi(s_t)$. More general formulations that include randomized decision making process, are not considered in this paper. Our smart device models seek to operate with the policy, $\pi(s)$, that maximizes over actions the expectation value of the total discounted reward, $\sum_{t=0}^\infty \gamma^t R_{a_t}(s_t,s_{t+1})$ over the Markov process, $P_a(\cdot,\cdot)$,  where $0<\gamma\leq 1$,  is the discount rate. There are numerous algorithms used for optimizing the policies. In our work we use the algorithms implemented in MDP Matlab toolbox \cite{MDP-package}.


\section{Models of Devices}
\label{sec:devices}

\begin{figure}[b]
 \begin{center}
 \includegraphics[width=0.45\textwidth]{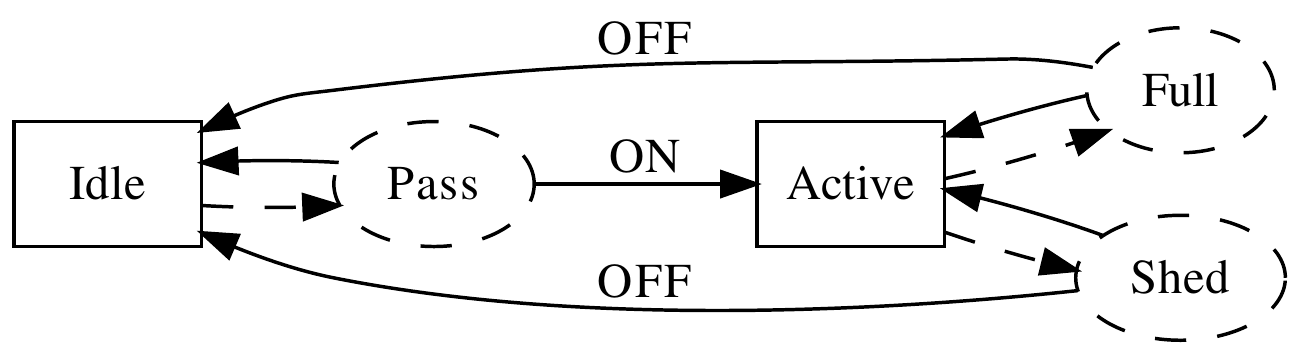}
 \caption{MDP diagram for the model of optional load. See text for explanations.}
 \label{fig:optional}
 \end{center}
\end{figure}

The specifics of our MDP setting are to be described below for four examples of loads. Note that these examples are meant to illustrate the power of the framework and its applicability to "smart grid" problems. In this first paper, we do not aim to make the examples realistic.  Instead, we focus on the qualitative features of the loads.
The states and actions associated with the devices are illustrated in the diagrams shown in Figs. \ref{fig:optional}-\ref{fig:storage}. For simplicity, we ignore  the external part of the state $s^{(e)}$ in these diagrams. Full diagrams can be produced by taking the Kronecker product of transition graphs associated with the device and the external factors. In our diagrams, the states are marked by squares and actions are marked by dashed circles. Transitions from states to actions and actions to states are marked by dashed and solid arrows, respectively.

\subsection{Optional Loads}
\label{subsec:OptLoads}

A smart device described by an ``optional load'' pattern can operate in two regimes, at full and limited capacity.  An example of such load is a light that can be automatically dimmed if the electricity price becomes too high (see Fig.~\ref{fig:optional}). To simplify the mathematical notations, we denote the states of the machine $s^{(m)}$ by $x$. The machine can be in either of the two states: $x=0$ and $x=1$, shown as $Idle$ and $Active$ in the diagram (\ref{fig:optional}) respectively. In the $x=0$ state the machine does not operate (the lights are off). In the $x=1$ state, the machine is active and the lights are shining at the full brightness, or are dimmed.  Actions of the device are $a_0 = \mbox{pass}$, $a_1=\mbox{full}$ or $a_2=\mbox{shed}$. The $a_0$ action represents the process of waiting for the external signal of switching on the device. If no external external signal (requesting switching on) appears,  the system returns to the $x=0$ state, otherwise it moves to the $x=1$ state. When the device is active (in the $x=1$ state), it has two options: operate at full capacity, corresponding to the action $a_1$, or shed the load (dim the lights), corresponding to action $a_2$. Turning the device on or off is an externality dependent on a human.  We assume that the external/human action is random,  with the probability of turning the device ON and turning the device OFF being $\rho_{ON}$ and $\rho_{OFF}$ respectively. (For simplicity, we assume that the OFF signal may arrive only by the end of the time interval.) Assuming additionally that the transition probabilities do not not depend on the device actions, we arrive to the following expression for the transition kernel:
\begin{eqnarray} \label{transition}
& P_{pass}(s,s') = T(c'|c)\left[\rho_{ON} \delta_{x',1} + (1-\rho_{ON})\delta_{x',0}\right], \\
& P_{full,shed}(s,s') = T(c'|c)\left[\rho_{OFF} \delta_{x',0} + (1-\rho_{OFF})\delta_{x',1}\right],
\end{eqnarray}
where $\delta_{x_1,x_2}$ is the Kronecker symbol: it is unity if $x_1=x_2$ and zero otherwise.

There is no reward associated with either outcome of the $a_0 = \mbox{pass}$ action,
however, the other two actions ($a_1$ and $a_2$) result in a reward consisting of two contributions.  First is the price paid to the
electricity provider, $E_{full,shed} c$,  where $c(t)$  is the cost of electricity (considered as a component of $s^{(e)}$) and $E_{full,shed}$ is the amount of energy consumed during the time interval which depends on whether the lights are fully on or dimmed. ( Here, $E_{full}>E_{shed}>0$ and both values do not depend on the resulting state of the device). Second,  the reward function accounts for a subjective level of comfort associated with the $a_{1,2}$ actions: $C_{full,shed}$. The discomfort of the light dimming is accounted by choosing $C_{full} > C_{shed}$. Summarizing,  the cumulative reward function in this model of the optional load becomes
\begin{eqnarray}
& R_{pass}(s,s') = 0, \\
& R_{full}(s,s') = C_{full} - E_{full} c, \\
& R_{shed}(s,s') = C_{shed} - E_{shed} c.
\end{eqnarray}

Obviously,  our model of optional loads is an oversimplification because there are a variety of additional effects which may also be important in practice, however, all these can be readily expressed within the MDP framework. For example,  one may need to limit the wear and tear on the device, thus encouraging (via a proper reward) minimization of switching. (To account for this effect would require splitting the $Active$ state in the model explained above into two states $Active-Full$ and $Active-Shed$.)

\subsection{Deferable Loads}
\label{subsec:DefLoads}

\begin{figure}
 \begin{center}
 \includegraphics[width=0.45\textwidth]{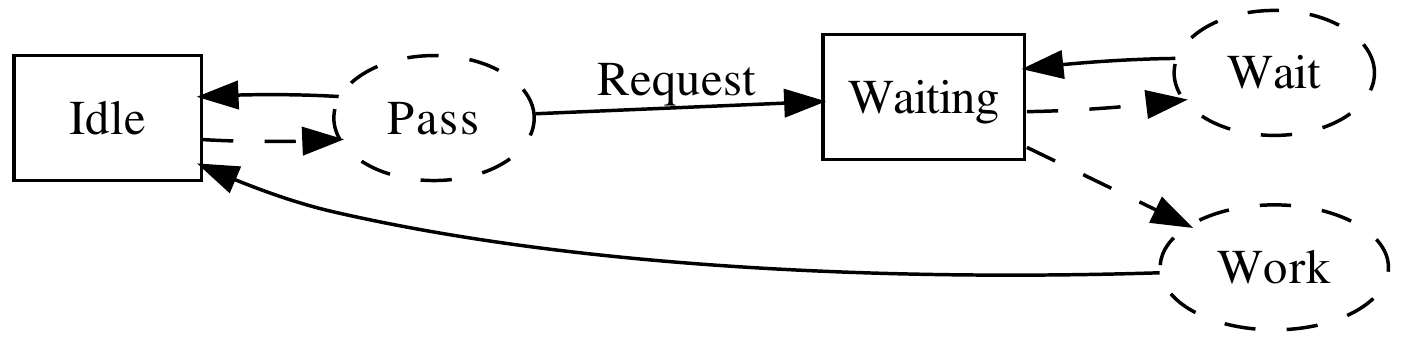}
 \caption{MDP diagram for the model of deferable load. See text for explanations.}
 \label{fig:deferable}
 \end{center}
\end{figure}

Our second example model is a deferable load,  i.e. a load whose operation can be delayed without causing a major consumer discomfort. Practical examples include dishwashing machines or some maintenance jobs like disk defragmentation on a computer. A simple model of such a device, shown in Fig. \ref{fig:deferable}, has two states: $x=0$ ($Idle$) when no work is required and $x=1$ ($Waiting$) when a job has been requested and the machine is waiting for the right moment (optimal in terms of the cost) to execute it. As in the previous model, the only action of the machine in the $Idle$ state is $a_0$ ($Pass$), however, in the $Waiting$ state, there are two possible actions: $a_1=Wait$ results in waiting for possible drop of the electricity price and $a_2=Work$ results in immediate execution of the job. The transition kernel for the model is
\begin{eqnarray}
& P_{Pass}(s,s') = T(c'|c)\left[\rho_{ON} \delta_{x',1} + (1-\rho_{ON})\delta_{x',0}\right], \\
& P_{Wait}(s,s') = T(c'|c)\delta_{x',1}, \\
& P_{Work}(s,s') = T(c'|c)\delta_{x',0},
\end{eqnarray}
where $\rho_{ON}$ is the probability of an exogeneous job request. In this model, there is no reward for choosing the $a_0=Pass$ action. The reward for the $a_2=Work$ action is equal to minus the price paid for the electricity, $R_{Work}(s,s') = - E*c$, and the reward for the $a_1=Wait$ action represents the level of discomfort associated with the delay, $R_{Wait} = C_{delay} < 0$. As in the model of optional loads, $E$ and $C_{delay}$ are constants parameters.


\subsection{Control Loads}
\label{subsec:ContrLoads}

\begin{figure}
 \begin{center}
 \includegraphics[width=0.45\textwidth]{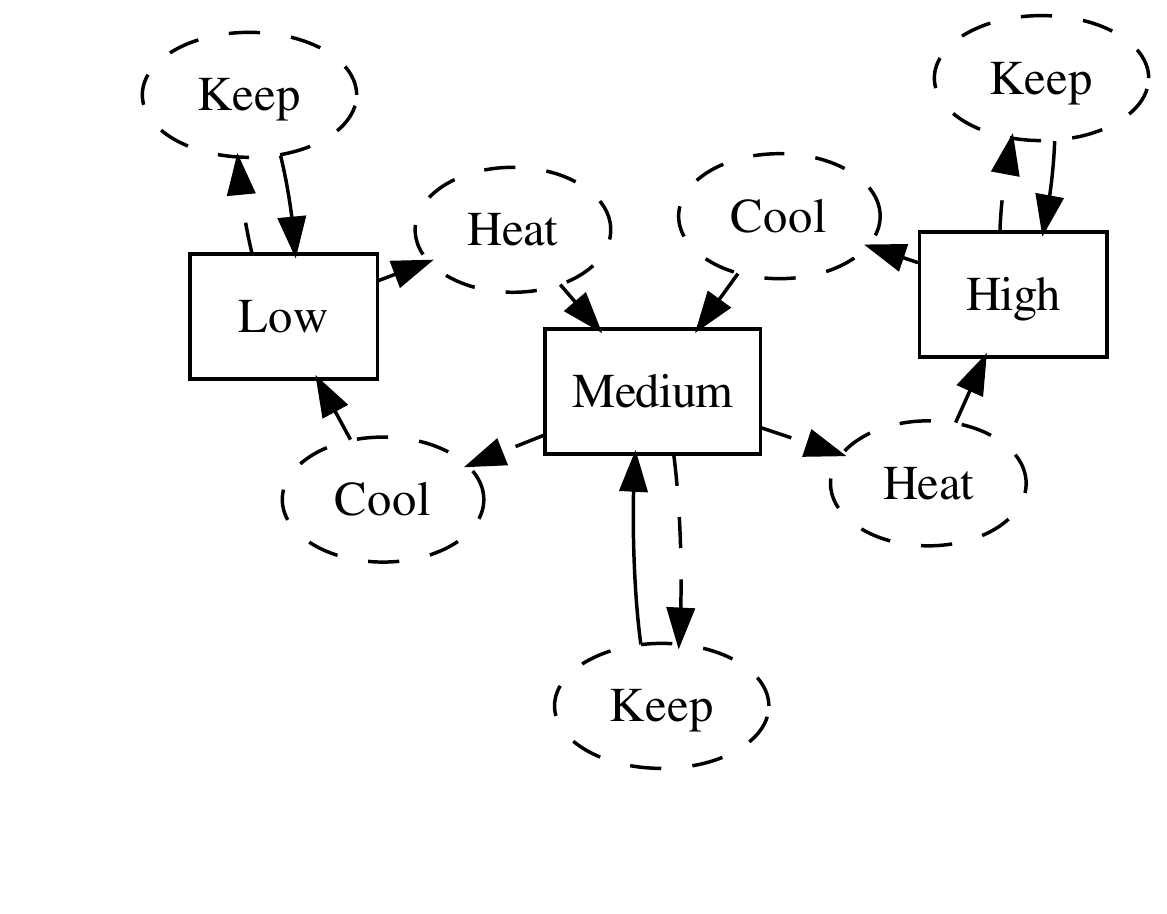}
 \caption{MDP diagram for the model of controllable load. See text for explanations.}
 \label{fig:control}
 \end{center}
\end{figure}

A very important class of devices that will likely play a key role in future demand response technologies are machines tasked to maintain a prescribed level of physical characteristics of some system. For example, thermostats are tasked with keeping the temperature in a building within acceptable bounds. Other examples of the control devices are water heaters, electric ovens, ventilation systems, CPU coolers etc.

In our enabling, proof-of-principle model of the control load, we consider a thermostat responsible for temperature control in a residential home. The state of the device is fully characterized by temperature which can take three possible values: $x=0,1,2$ corresponding to $Low, Medium, High$ temperatures, respectively.  Each temperature is assumed to be operationally acceptable. For simplicity, we assume that the thermostat uses an electric heater to modify the temperature (i.e. the outside temperature is low). The device can choose between the following three actions.  $a_0 = Cool$ leaves the heater idle for the forthcoming interval. Since there is some base consumption associated with the thermostat operation we assume that $E_{Cool}>0$. The next action, $a_1=Keep$, maintains the temperature at the current level and requires some energy for heater operation: $E_{Keep}>E_{cool}>0$. Finally, $a_2=Heat$ corresponds to intensive heating that raises the temperature and requires the largest amount of energy $E_{Heat}$, and $E_{Heat}>E_{Keep}>E_{Cool}=0$. Our thermostat state diagram, shown in Fig.~(\ref{fig:control}), assumes  that the dynamics of the thermostat are deterministic, and the resulting state depends only on the action chosen. The transition probabilities of the thermostat MDP is
\begin{eqnarray}
& P_{Heat}(s,s') = T(c'|c)\delta_{x',x+1}, \\
& P_{Keep}(s,s') = T(c'|c)\delta_{x',x}, \\
& P_{Cool}(s,s') = T(c'|c)\delta_{x',x-1}.
\end{eqnarray}
Assuming that all levels of temperature are equally comfortable, the reward function depends only on the price and energy consumption associated with the action,
\begin{eqnarray}
R_{Cool,Keep,Heat}(s,s') = - c E_{Cool,Keep,Heat}.
\label{reward_control}
\end{eqnarray}
Our basic model can be generalized to account for different comfort levels of different states, the possibility for the owner to override an action, variations of the outside temperature, etc.

\subsection{Storage loads}
\label{subsec:Storage}

\begin{figure}
 \begin{center}
 \includegraphics[width=0.45\textwidth]{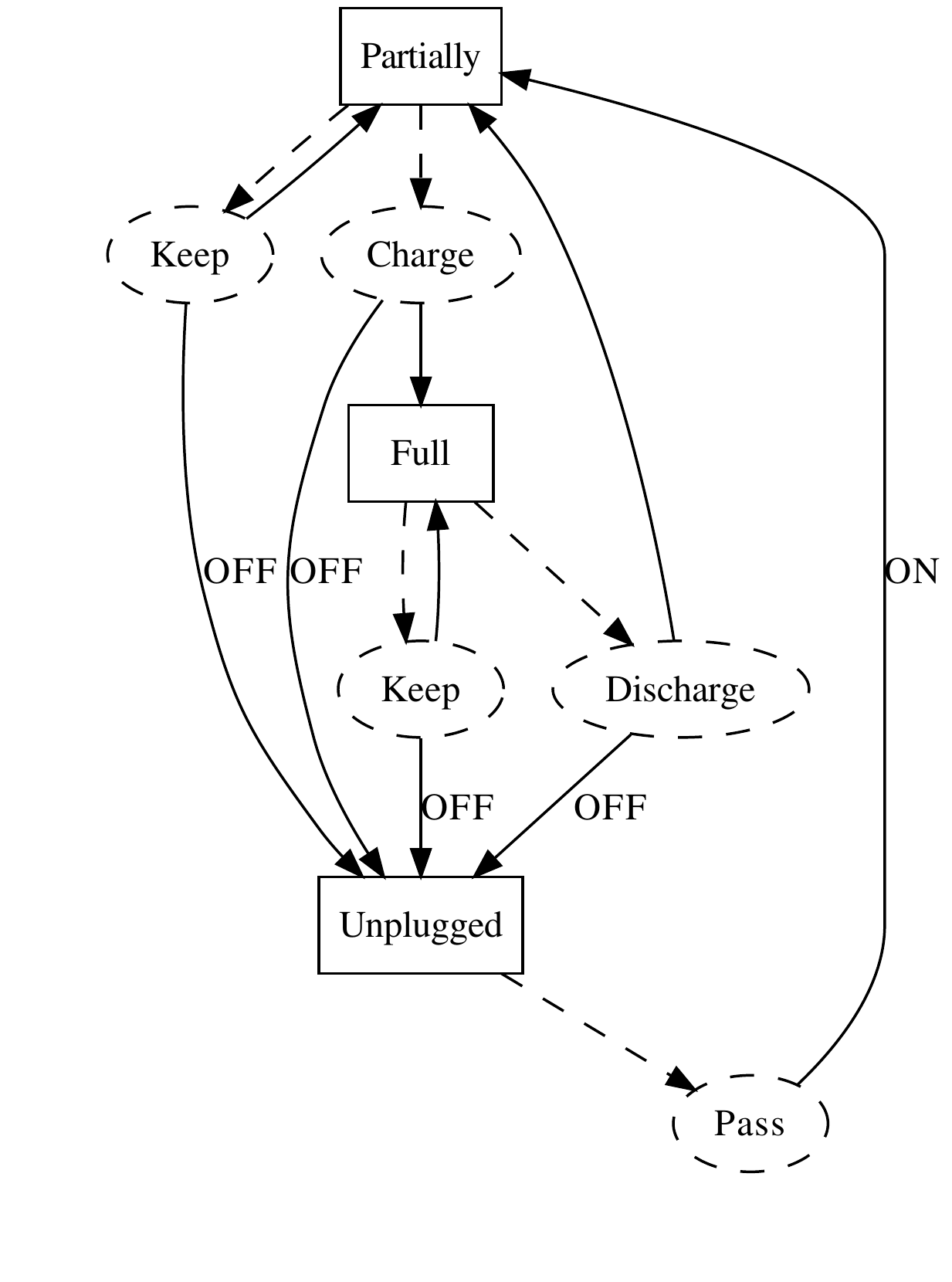}
 \caption{MDP diagram for the model of storage. See text for explanations.}
 \label{fig:storage}
 \end{center}
\end{figure}

The number of devices with rechargeable batteries is expected to increase dramatically in the coming years. Currently, these are mostly laptops, uninterruptable power supplies, etc. In addition, a significant number of large-scale batteries will be added to the grid most likely via the anticipated Plug-in Hybrid Electric Vehicles (PHEV) potentially enabled with Vehicle-to-Grid (V2G) capability.  Storage devices, illustrated with the MDP in Fig.~(\ref{fig:storage}),  share some similarity with the controlled loads discussed in the previous Subsection,  but they are also different in two aspects. First, users/owners wants their devices to be charged which leads to a level of discomfort if the devices are not fully charged. Second, and probably most significantly,  storage devices such as PHEVs are disconnected from the grid when in use. Having PHEVs in mind,  we propose the following model of (mobile) storage. The system can be in either of the three states, the $x=0=Unplugged$ state (which is similar to the Idle state in the models of Optional and Deferable loads discussed above),  the $x=1=Partially$ state where the storage is partially charged, and the $x=2=Full$ state where the device is fully charged.

The four available actions are: $a_0=Pass$ when the device is in the unplugged state, the $a_1 = Keep$ action possible when the initial state is $x=1=Partially$ or $x=2=Full$, the $a_2=Charge$ action available from the $x=1=Partially$ state which transitions to the $x=2=Full$ state, and, finally, the $a_3=Discharge$ action, that is an inverse of the $a_2$ one, available from the $x=2=Full$ state resulting in the $x=1=State$. Except for $a_0=Pass$, all these actions can be interrupted by transitioning at the end of the time interval to the $x=0=Unplugged$ state. As in previous sections, we assume that the unplugging happens at the end of a time interval. Assuming the device can be unplugged with the probability $\rho_{OFF}$ and that it can be reconnected to the grid with the probability $\rho_{ON}$, we arrive at the following expressions for the transition probability:
\begin{eqnarray}
& P_{Pass}(s,s') = T(c'|c)\left[\rho_{ON}\delta_{x',1} + (1-\rho_{ON})\delta_{x',0}\right], \\
& P_{Keep}(s,s') = T(c'|c)\left[\rho_{OFF}\delta_{x',0} + (1-\rho_{OFF})\delta_{x,x'}\right], \\
& P_{Charge}(s,s') = T(c'|c)\left[\rho_{OFF}\delta_{x',0} + (1-\rho_{OFF})\delta_{x',2}\right],\\
& P_{Discharge}(s,s') = T(c'|c)\left[\rho_{OFF}\delta_{x',0} + (1-\rho_{OFF})\delta_{x',1}\right].
\end{eqnarray}
The reward function accounts for the following effects.  First, the $a_1=Keep$ action has the cost associated with keeping the battery charged, $E_{Keep}(x)$, naturally dependent on the state, $E_{Keep}(2)>E_{Keep}(1)>E_{Keep}(0)=0$. Second, the $a_2=Charge$ action requires $E_{Charge}$ of energy while the $a_3=Discharge$ action generates the $E_{Discharge} < 0$ of energy,  both nonzero only if the resulting state is not the $x=0=Unplugged$.  Therefore, all the ``active" actions, $Keep, Charge, Discharge$, contribute the reward function in accordance with the energy price, $c' E_{\dots}$. Finally, we also assign an additional negative reward, $C_{Unplug} <0 $, accounting for the discomfort (to the human) associated with being in the $x=0=Unplugged$ state. The resulting reward function is
\begin{eqnarray}
& R_{Pass}(s,s') = 0,\\
& R_{Keep}(s,s') = C_{Unplug} \delta_{x',0}\delta_{x,1} - c E_{Keep}(x),\\
& R_{Charge}(s,s') =-c E_{Charge}, \\
& R_{Discharge}(s,s') = C_{Unplug}\delta_{x',0} - c E_{Discharge} .
\end{eqnarray}

\section{Simulations}
\label{sec:Simulations}

In order to illustrate the capabilities of the proposed framework, we consider a simple model of the control load, describing a smart thermostat, characterized by $N_T=10$ levels of the temperature parameter $T$. At every moment of time the thermostat can choose to raise, lower or keep the same temperature.  The raise and lower options are not available at the highest and lowest possible temperatures, respectively. The energy consumption associated with the actions is given by  $E_{Keep} = 1.0$, $E_{Cool} = 0.1$ and $E_{Heat} = 2.1$, respectively, in some normalized energy units. This choice of energies discourages the system from switching the heater too often: although the combinations $Heat + Cool$ and $Keep + Keep$ lead to the same temperature levels, the latter action is preferable as it consumes less energy.

\begin{figure}
 \begin{center}
 \includegraphics[width=0.45\textwidth]{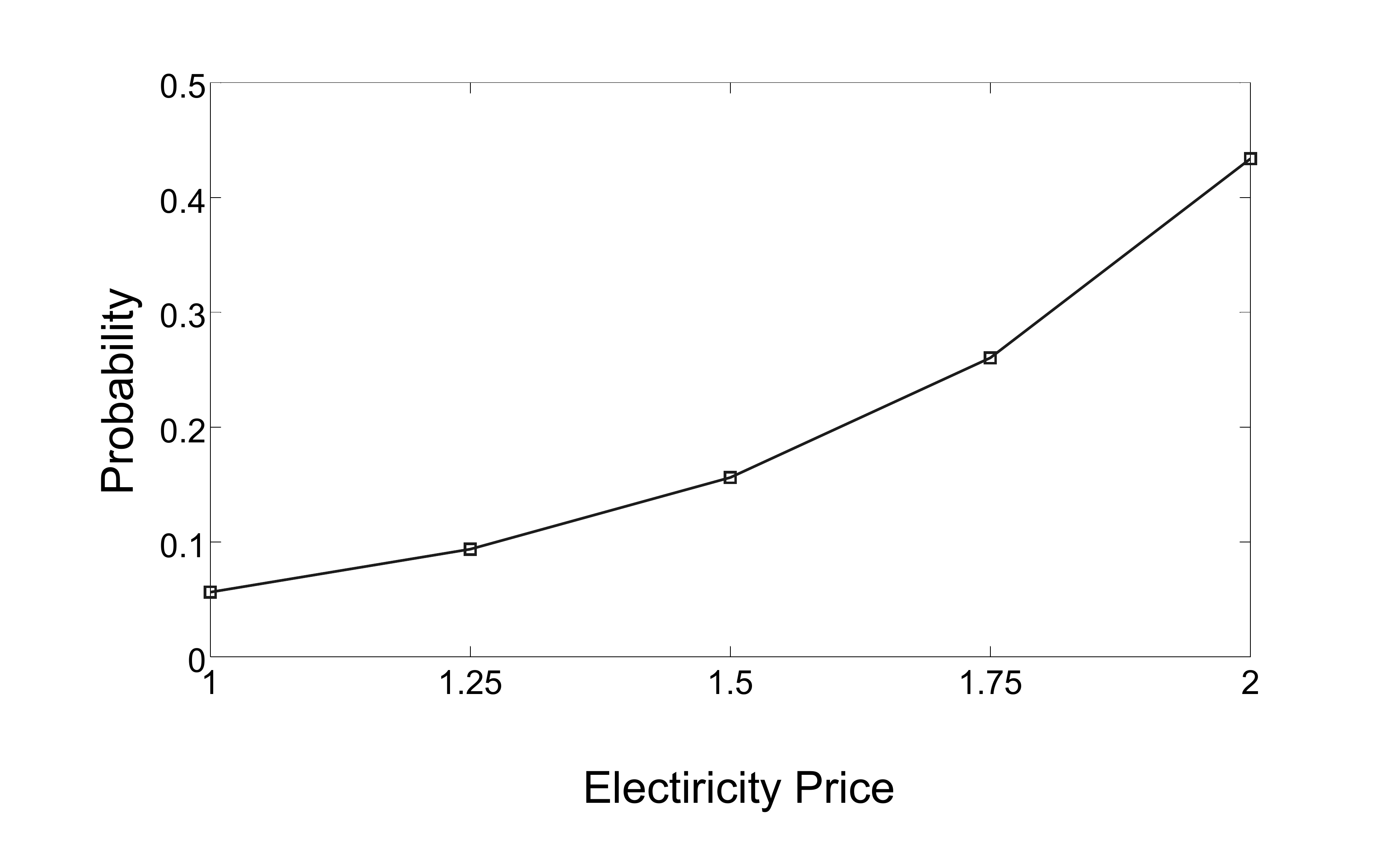}
 \caption{Probability distribution of electricity price in the model example.}
 \label{fig:price-dist}
 \end{center}
\end{figure}

Variations in price are modeled by a Markov chain of $N_P=5$ equidistant levels with the minimum and maximum corresponding to $1.0$ and $2.0$ price units, respectively. At each time interval, the price either increases with probability $T(c+1|c) = 0.5$ by $1$ level, decreases with probability $T(c-1|c) = 0.3$ by $1$ level, or stays the same. The resulting stationary probability distribution $p(c)$ is shown in the Figure \ref{fig:price-dist}. It is skewed towards the higher price, mimicking the effect of intermittent renewable generators that occasionally provide excess power to the grid, thus leading to rapid dips in the price.
The reward function (\ref{reward_control}) is fully determined by the total cost of energy consumed by the thermostat within the given time-interval. Our MDP model imposes upper and lower bounds on the temperature, and we assume that there is no additional discomfort associated with the variations of temperature between these bounds, i.e. all of the $N_T$ temperature levels are equally comfortable for the consumer.

\begin{figure}
 \begin{center}
 \includegraphics[width=0.45\textwidth]{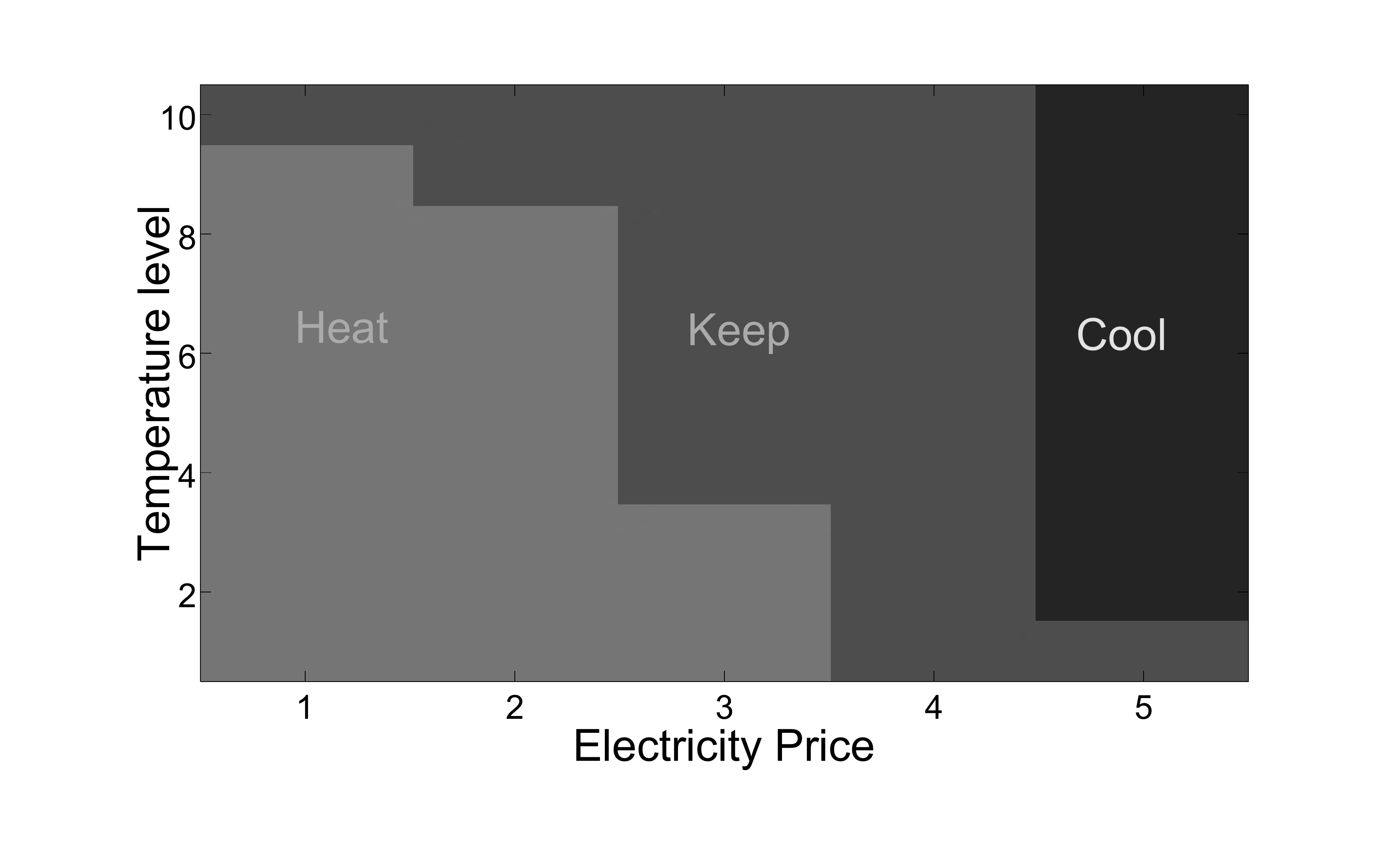}
 \caption{Visualization of the policy found as a result of optimization.}
 \label{fig:policy}
 \end{center}
\end{figure}

\begin{figure}
 \begin{center}
 \includegraphics[width=0.45\textwidth]{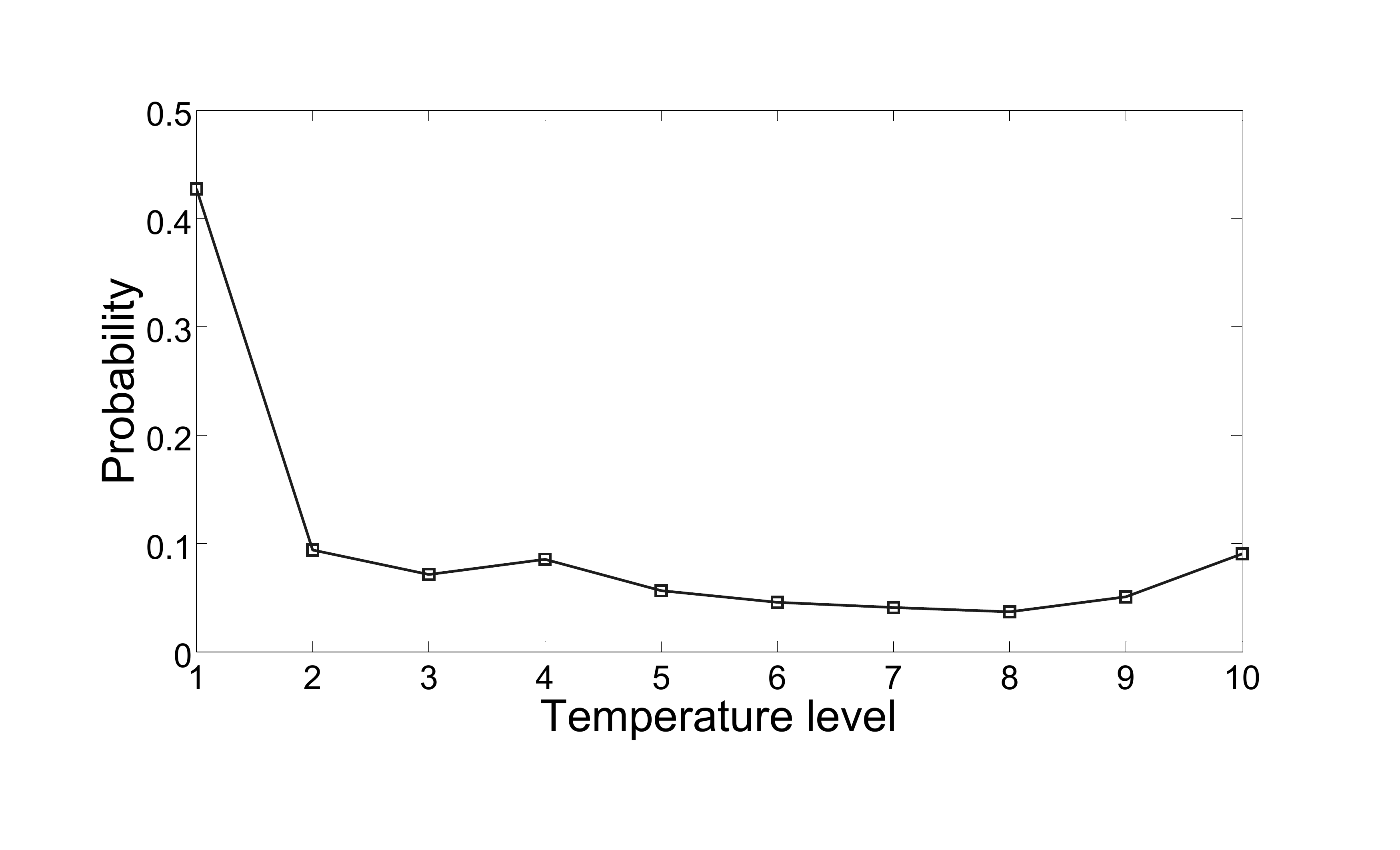}
 \caption{Probability distribution of temperature levels observed at the optimal policy.}
 \label{fig:temperature}
 \end{center}
\end{figure}

This system was analyzed with the Matlab MDP package \cite{MDP-package} where we used different algorithms to verify the stability of the results. The resulting optimal policy (for the range of parameters tested) is illustrated in \ref{fig:policy}. As expected, the thermostat chooses the $Heat$ action when the price is low and decides to $Cool$ when the price is high; a set of actions that lead to the skewed probability distribution of temperatures shown in Figure \ref{fig:temperature}. One finds that the thermostat spends most of the time performing $Keep$ in the low temperature state waiting for the price to drop.

\begin{figure}
 \begin{center}
 \includegraphics[width=0.45\textwidth]{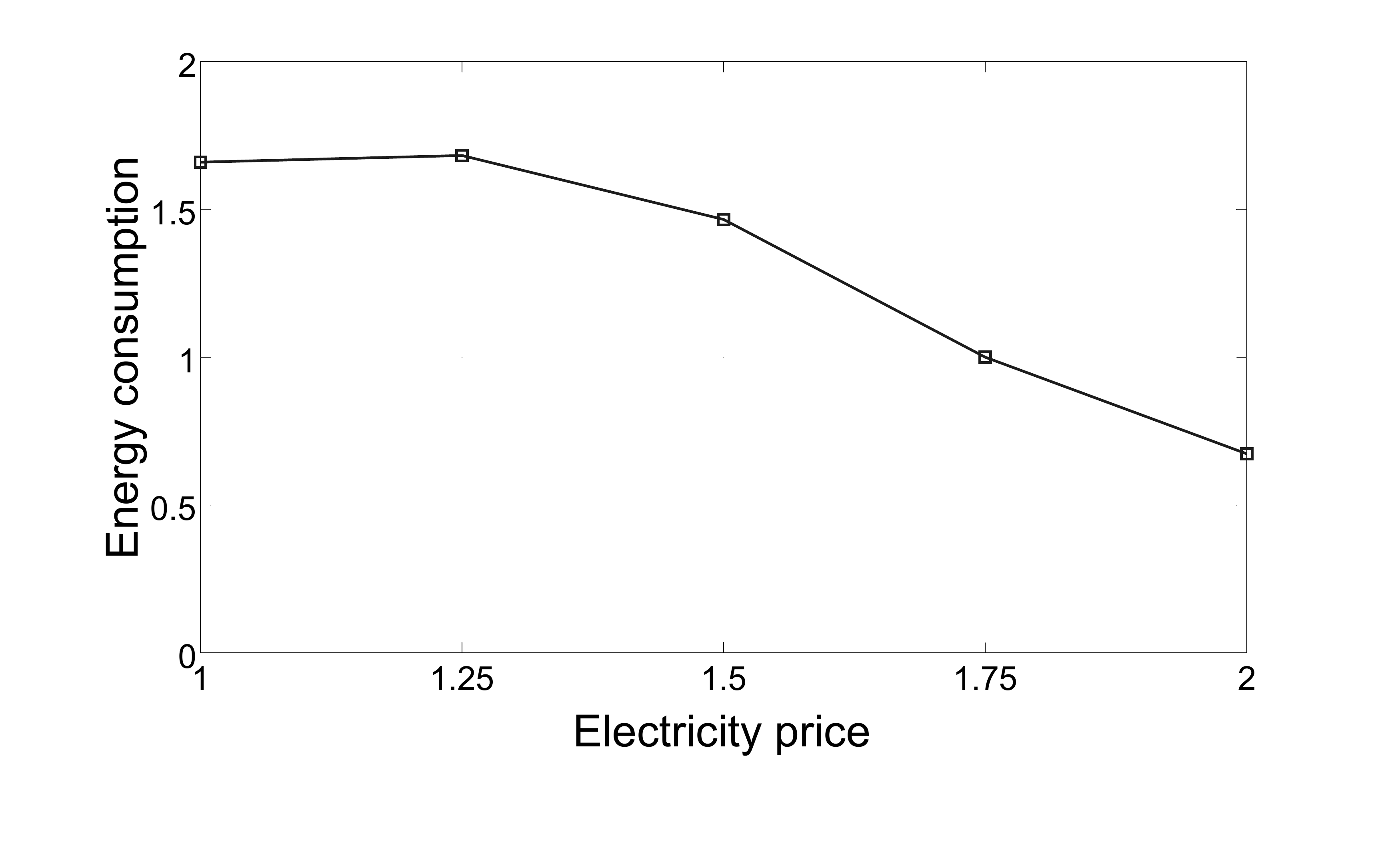}
 \caption{Demand of the smart thermostats.}
 \label{fig:elasticity}
 \end{center}
\end{figure}

Perhaps, the most interesting feature of the MDP model is the relation between consumption and price. We define the expected demand as the average energy demand for a given price
\begin{equation}
 \langle E|c\rangle  = \frac{\sum_{x} E_{\pi(x,c)} P_{st}(x,c) }{\sum_{x} P_{st}(x,c)},
 \label{elacticity}
\end{equation}
where $P_{st}(x,c)$ is the stationary joint distribution function of the temperature and price at the optimal strategy. Dependence of the consumption on the price for our choice of the parameters is shown in Figure \ref{fig:elasticity}, thus illustrating that variations in price indeed produce demand response. An interesting feature is that the demand curve is not monotonic. At low temperatures, the energy consumption shows a slight increase with the price; a surprising behavior related to saturation of the demand. When the electricity price decreases gradually from high to low levels, there is a high probability that the thermostat will reach the highest level of temperature before the price reaches the lowest level. In this case, the demand will be lower at the smallest price levels as there will be no unsatisfied demand left in the system to capitalize on the lowest price. From the economic viewpoint, it is important to note that this non-monotonicity of the demand curve reflects the adaptive nature of the MDP algorithm: the smart devices adjust to fluctuations in price, thus making it more difficult for the electricity providers to exploit the non-monotonic demand curve for making profit.

Another interesting result found in our simulations is an increase in average consumption of the smart (policy optimized) thermostat when compared to its non-smart counterpart, where the latter is defined as the one ignoring price fluctuations and sticking to the $Keep$ action. For the set of parameters chosen in the test case, we observed that the average level of consumption in the optimal case is $1.03$, i.e. it is $3\%$ higher than in the naive strategy, an effect associated with the additional penalty (in energy) imposed on the $Heat$ and $Cool$ actions.

It is also instructive to evaluate savings of the consumer. The average value of the reward associated with the optimal policy is equal to $-1.6722$, which should be compared with the reward of $-1.73$ generated by its non-smart counterpart. Since the reward reflects the customer's cost of electricity, we conclude that the customer saves about $3\%$ on the electricity costs associated with the thermostat.  The lower total energy costs for higher energy consumption was also seen in a related ``smart-device'' demonstration project\cite{Olympic}.

Note,  that the quantitative conclusions drawn and numbers presented above were meant to illustrate the questions the MDP approach can resolve. The conclusions and the numbers do not represent any real device as the parameters used were not justified by actual data.

\section{Discussions, Conclusions and Path Forward}
\label{sec:Path}

To conclude, we have presented a novel modeling framework to analyze future demand response technologies. The main novel aspect of our approach lies in the capability of the framework to describe behavior of the smart devices under varying/fluctuating electricity prices. To achieve this goal, we modeled the devices as rational agents which seek to maximize a predefined reward function associated with its actions. In general, the reward function includes the price paid for the electricity consumption and the level of owner discomfort associated with the choices made by the device. At the mathematical level, the system can be described via Markov Decision Processes that have been extensively studied over the last 50 years. Utilizing the MDP approach, we showed that a great variety of practical devices can be described within the same framework by simply changing the set of device states, actions and reward functions. Specifically, we  identified four main device categories and proposed simple MDP models for each of them. These four categories include optional loads (like light dimming), deferrable loads (like dishwashing), control loads (thermostats and ventilation systems), and finally storage loads (charging of batteries).

To illustrate the approach we experimented with a simple model of a smart heating thermostat. The MDP-optimized policy of the thermostat followed the expected pattern: it chooses to not heat or keep the temperature stationary at high prices and prefers to heat when the price is low. This policy resulted in $3\%$ of savings in the price paid for electricity, but at the same time led to the total of $3\%$ increase in the consumption level due to the energy costs associated with the thermostat actions. The resulting demand curve showed a noticeable amount of elasticity, thus meeting the main objective of the demand response technology.

There are many relevant aspects of the model that we did not discuss in the manuscript. We briefly list some of these and future research challenges and direction.
\begin{itemize}
\item \emph{Learning algorithms}. In our model we assumed that smart devices have an accurate model of stochastic dynamics for external factors (such as price for electricity), and use this model to find the optimal policy. In reality, however, this model is not known ab initio and has to be learned from the observations. Moreover, one can expect that the dynamics of external factors will be highly non-stationary (i.e. the transition matrix $T(s^{(e)}_{t+1}|s^{(e)}_{t+1})$ will have an explicit dependence on time).  Therefore, the optimal policy has to be constantly adapted to the varying dynamics of the external factors. Of a special practical interest is the generalization of the framework to almost periodic processes, reflecting natural daily/weekly/yearly cycles in the electricity consumption.

\item \emph{Price-setting policies}. We did not discuss the price setting policies above, assuming that the policies are given/pre-defined. However, the electricity providers might adjust their policies to consumer response. As the electricity providers pursue their own goals, this setting essentially becomes game-theoretic and as such it requires more sophisticated approaches for analysis. Another extension of the model is to introduce auction-based price-setting schemes, such as in the Olympic Peninsula project \cite{Olympic}. This setting can be naturally incorporated in the same framework, although the modification may require simultaneous modeling of multiple (ensemble of) devices.

\item \emph{Time delays}. Another aspect of the real world not incorporated in our analysis concerns the separation of the time scales associated with operations of the device and intervals of the price variations. Multiple time-scale can be naturally incorporated in the framework by introducing additional states of the device. These modifications will certainly affect final answer for the optimal policy, and the resulting demand curve. However, accurate characterization of the multi-scale behavior will be a challenging task, requiring analysis of nonlinear response functions and dynamical description of the underlying non-Markovian processes.
\end{itemize}

\section{Acknowledgements}

We are thankful to the participants of the ``Optimization and Control for Smart Grids" LDRD DR project at Los Alamos and Smart Grid Seminar Series at CNLS/LANL for multiple fruitful discussions.

\bibliographystyle{IEEETran}
\bibliography{demand_response}

\end{document}